\documentclass[aps,prc,twocolumn,showkeys,showpacs,groupedaddress,superscriptaddress]{revtex4}
\usepackage{amssymb}
\usepackage{graphicx}% Include figure files
\usepackage{color}

\usepackage{amsmath}
\newcommand{\req}[1]{(\ref{#1})}
\newcommand{\bel}[1]{\begin{equation}\label{#1}}
\newcommand{\belar}[1]{\begin{eqnarray}\label{#1}} 

\begin{document}
%\selectlanguage{russian}
%\title{Description of the reactions $^{36}$S + $^{238}$U and $^{64}$Ni + $^{238}$U within the two-stage fusion-fission model}
\title{Formation of superheavy nuclei  in $^{36}$S + $^{238}$U and $^{64}$Ni + $^{238}$U reactions}
\author{V. L. Litnevsky}
\email{vlad.lit@bk.ru}
\affiliation{Omsk State Transport University, 644046 Omsk, Russia}

\author{ F. A. Ivanyuk}
\email{ivanyuk@kinr.kiev.ua}
\affiliation{Institute for Nuclear Research, 03028 Kiev, Ukraine}

\author{G. I. Kosenko}
\email{kosenkophys@gmail.com}
\affiliation{Omsk Tank Automotive Engineering Institute, 644098 Omsk, Russia}

\author{S. Chiba}
\email{chiba.satoshi@nr.titech.ac.jp}
\affiliation{Tokyo Institute of Technology, 152-8550 Tokyo, Japan}

\date{today}

\begin{abstract}
We describe the capture, fusion, fission and evaporation residue formation cross sections of superheavy nuclei within the proposed earlier two stages dynamical model. The approaching of the projectile nucleus to the target nucleus is described in the first stage of the model. On the second stage, the evolution of the system formed after the touching of the projectile and target nuclei is considered. The evolution of the system on both stages is described by Langevin equations. The transport coefficients of these equations are calculated within the microscopic linear response theory. The mutual orientation of the colliding ions, the tunneling through the Coulomb barrier in the entrance channel and the shell effects in the potential energy on both stages of the calculations are taking into account. The obtained results are compared with the available experimental data and other theoretical predictions.
\end{abstract}


\maketitle

\section{Introduction}

One of the most interesting and intensively developing branches of nuclear physics is the synthesis of superheavy elements. Unfortunately, reactions in which two spherical in the ground-state nuclei collide with each other (the cold fusion reactions) have exhausted themselves. The further studies of superheavy elements involve the hot fusion reactions in which a spherical projectile nucleus interacts with a heavy deformed target nucleus.

The theoretical models of such reactions have to take into account the shell structure of colliding nuclei (in order to reproduce the non-spherical shape of the target nucleus in the ground state). Also, the initial orientation of the target nucleus relative to the line, connecting the centers of mass of colliding nuclei, should be taken into account. Finally,  the possibility of deformation of the nuclei during the collision must be taken into account.

In the present paper, the reactions $^{36}$S+$^{238}$U $\to^{274}$Hs  and ${\rm ^{64}Ni+^{238}U}$ $\to{\rm ^{302}120}$ are investigated using the two stages dynamic stochastic model \cite{Shen:02,koivpa}. These reactions differ significantly from each other by the ratio of the masses and charges of the colliding nuclei. It is known well, that with increasing mass asymmetry of colliding nuclei, there is a noticeable increase in the compound nuclei formation cross-section. Thus, the comparison of calculated results for the considered reactions with the experimental data allows us to judge the ability to apply the developed model to the analysis of a wider range of the SHE formation reactions.

\section{The model}

In the used here model the time evolution of the system of two colliding ions is described by the Langevin equations \cite{Abe:96,Marten:92} for the shape degrees of freedom.
At both stages of calculations we use the shape parameterizations based on the Cassini ovaloids \cite{Pashkevich:71}.

The deformation energy $E_{\rm def}^{(t)}$ and $E_{\rm def}^{(p)}$ of colliding ions and $E_{\rm def}$ of the combined system are defined with\-in the macroscopic-microscopic method \cite{swiat}. The shell and the pairing corrections to the liquid drop energy are calculated by the approach, proposed by Strutinsky \cite{Strutinsky:67,Brack:72}.

The evolution of collective coordinates, describing the state of the system (two separated ions in the entrance channel or the compact nucleus formed after the touching of ions)
is described in terms of the Langevin equations
\cite{Abe:96,Marten:92}, namely,
\begin{eqnarray}
\dot {q}_\beta&=&\mu_{\beta \nu}p_\nu , \nonumber\\
\dot {p}_\beta&=&-\frac{1}{2}p_\nu p_\eta\frac{\partial \mu_{\nu \eta}}{\partial q_\beta}+K_\beta-\gamma_{\beta \nu}\mu_{\nu \eta}p_\eta+\theta _{\beta \nu}\xi_\nu.
\label{eq:UL}
\end{eqnarray}
Here ${q}_\beta$ are the deformation parameters and
a convention of
summation
over repeated indices $\nu$, $\eta$ is
used.
%meant.
The quantity
$\gamma_{\beta\nu}$ is the tensor of friction coefficients and
$\mu_{\beta\nu}$ is the tensor inverse to the mass tensor
$m_{\beta\nu}$,

At both stages of calculations the friction $\gamma_{\beta\nu}$
and inertia $m_{\beta\nu}$ tensors are calculated within the linear response
approach and local harmonic approximation \cite{Hofmann:97,Hofmann:08}. In
this approach many quantum effects such as shell and pairing
effects, and the dependence of the collisional width of the single particle
states on the excitation energy, are taken into account.
For slow collective motion the tensors of friction $\gamma_{\beta\nu}$ and inertia $m_{\beta\nu}$ can be expressed in terms of first and second derivatives of the Fourier transform $\chi_{\beta\nu}(\omega)$  of the response function,
\begin{eqnarray}\label{chiom}
\chi_{\beta\nu}(\omega)&=&\sum_{kj}\xi_{kj}^2
\frac{n_k^T-n_j^T}{\hbar\omega -E_{kj}^-+i\Gamma_{kj}}F^{\beta}_{kj}F^{\nu}_{jk}
\nonumber\\
&+&\sum_{kj}\eta_{kj}^2 \frac{n_k^T+n_j^T-1}{\hbar\omega-E_{kj}^++i\Gamma_{kj}}F^{\beta}_{kj}F^{\nu}_{jk}\,,
\end{eqnarray}
\begin{equation}\label{zerofric}
\gamma_{\beta\nu}=-i\frac{\partial\chi_{\beta\nu}(\omega)}{\partial\omega}\Bigr\arrowvert_{\omega=0}\,,
\,
m_{\beta\nu}=\frac{1}{2}\frac{\partial^2\chi_{\beta\nu}(\omega)}{\partial\omega^2}\Bigr\arrowvert_{\omega=0}.
\end{equation}
The precise expressions for the friction are:
\begin{eqnarray}\label{friczero}
\gamma_{\beta\nu}=2\hbar{\sum_{kj}}(n_j^T-n_k^T)\xi_{kj}^2%^{\pr}
\frac{E_{kj}^-\Gamma_{kj}}{[(E_{kj}^-)^2+\Gamma_{kj}^2]^2}F^{\beta}_{kj}F^{\nu}_{jk}
\nonumber\\
+2\sum_{kj} (1-n_k^T-n_j^T)\eta_{kj}^2\frac{E_{kj}^+\Gamma_{kj}}
{[(E_{kj}^+)^2+\Gamma_{kj}^2]^2}F^{\beta}_{kj}F^{\nu}_{jk}
\end{eqnarray}
and
\begin{eqnarray}\label{masszero}
m_{\beta\nu}=\hbar^2{\sum_{kl}}(n_j^T-n_k^T)\xi_{kj}^2
\frac{E_{kj}^-((E_{kj}^-)^2-3\Gamma_{kj}^2)}
{[(E_{kj}^-)^2+\Gamma_{kj}^2]^3}F^{\beta}_{kj}F^{\nu}_{jk}\nonumber\\
+\hbar^2\sum_{kj} (1-n_k^T-n_j^T)\eta_{kj}^2
\frac{E_{kj}^+((E_{kj}^+)^2-3\Gamma_{kj}^2)}
{[(E_{kj}^+)^2+\Gamma_{kj}^2]^3}F^{\beta}_{kj}F^{\nu}_{jk}.
\end{eqnarray}
Here $E_k, E_j$ are the energies of quasiparticle states in
BCS-approximation,   $E_{kj}^-\equiv E_{k}-E_{j},\, E_{kj}^+\equiv
E_{k}+E_{j}$,  $n^T_k\equiv 1/(1+\exp(E_k/T))$,
$\eta_{kj}=u_k\upsilon_j+u_j\upsilon_k,\,\xi_{kj}=u_ku_j-\upsilon_k\upsilon_j$, $u_k,\upsilon_k$ are the coefficients of Bogoliubov-Valatin
transformation.  The operator
$\hat{F}_{\beta}$, which appears in
(\ref{friczero})--(\ref{masszero}), is the derivative of the
single-particle Hamiltonian  with respect to the
deformation parameter $q_{\beta}$. The quantity $\Gamma_{kj}$  is
the average width of the two-quasiparticle states,
$\Gamma_{kj}=(\Gamma(E_k, \Delta, T)+\Gamma(E_j, \Delta, T))/2$.
The calculation of  $\Gamma_{kj}$ for the system with pairing is
explained in detail in \cite{ivahof}. One of us (F.I.) apologize very much for the misprints in expressions for $\gamma_{\beta\nu}$ and $m_{\beta\nu}$, given in \cite{Ivanyuk:1999}.

The $K_\beta$ in \req{eq:UL} is the component of conservative force $\vec K=-\bigtriangledown F$, where $F=V_{pot}-aT^2$ is the free energy of the system, $V_{pot}$ -- its potential (deformation) energy, $a$ is the level density parameter \cite{mebel:92} and the temperature $T$ of system  is related to the internal (dissipated) energy by
the Fermi-gas formula $T=\sqrt {E_{\rm dis}/a}$.

Friction provides the dissipation of collective motion energy into internal energy. The fluctuations in the system are described by the random force $\theta_{\beta\nu}\xi_\nu$.
Here $\xi_\nu$ is a random number with the following properties
\begin{eqnarray}
<\xi_\nu>&=&0,\nonumber\\
<\xi_\beta(t_1)\xi_\nu(t_2)>&=&2\delta_{\beta\nu}\delta(t_1-t_2).
\end{eqnarray}

The magnitude of the random force $\theta_{\beta\nu}$ is
expressed in terms of diffusion tensor $D_{\beta\nu}$,
$D_{\beta\nu}=\theta_{\beta\eta}\theta_{\eta\nu}$, which is
related to the friction tensor $\gamma_{\beta\nu}$ via the modified
Einstein relation $D_{\beta\nu}=T^*\gamma_{\beta\nu}$, where $T^*$ is the effective temperature \cite{hofkid},
\begin{equation}\label{T*}
 T^*=\frac{\hbar\varpi}{2}\coth{\frac{\hbar\varpi}{2T}} \,\ .
\end{equation}
 The parameter $\varpi
$ is the local frequency of collective motion \cite{hofkid}. The minimum of
$T^{\ast }$ is given by $\hbar \varpi /2$.

The total energy of the system is fixed at the initial stage,
\bel{etot}
E_{\rm tot}=E_{\rm gs}^{(t)}+E_{\rm gs}^{(p)}+E_{\rm c.m.}.
\end{equation}
Here $E_{\rm gs}^{(t)}, E_{\rm gs}^{(p)}$ are the ground state energies of the target and projectile, $E_{\rm c.m.}=E_{\rm lab}A_{\rm p}/(A_{\rm p}+A_{\rm t})$ is the energy of relative motion of target and projectile, calculated in the center-of-mass system, $A_{\rm p}$ and $A_{\rm t}$ are, correspondingly, the mass numbers of the target and projectile.
By introducing the $Q$-value of reaction
\bel{qval}
Q\equiv E_{\rm gs}^{(t)}+E_{\rm gs}^{(p)}-E_{\rm gs}^{(t+p)}
\end{equation}
the total energy can be written as
\bel{etota}
E_{\rm tot}=E_{\rm gs}^{(t+p)}+E_x \qquad{\rm with}\qquad E_x=E_{\rm c.m.}+Q
\end{equation}
The $E_x$ is the excitation energy of the system above the ground state of compound nucleus formed after fusion of target and projectile. The $E_x$ is fixed by the initial conditions and does not depend on time.
The calculations in the present work were carried out for a few values of $E_x$ mentioned below.

Some terms of the equation (\ref{eq:UL}) should be determined twice, ones for the first, and ones for the second stage of calculations. Such terms we will denote by the upper indexes  ($I$) and ($II$), respectively.
%%%%%%%%%%%%%%%%%%%%%%%%%%%%%%%%%%%%%%%%%%%%%%%%%%%%%%%%%%%%%%%%%%%
\subsection{The entrance channel}
\begin{figure}[htp]
\includegraphics[width=0.4\textwidth]{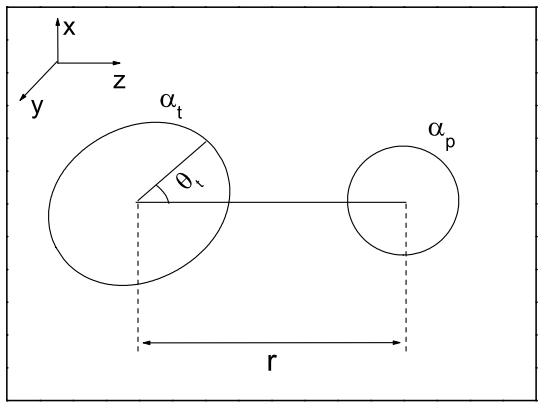}
\caption{Collective coordinates of the system, which consist of two separated nuclei. Shape of the system is determined by four parameters, namely, by the distance $r$ between the centers of mass of the colliding nuclei, by the deformation parameters of target ($\alpha_t$) and projectile ($\alpha_p$) nuclei, and by the orientational parameter $\theta_t$ which is an angle between the symmetrical axis of the deformed in the ground state target nucleus and the line connecting centers of mass of the colliding nuclei.}
\label{4param}
\end{figure}

%%%%%%%%%%%%%%%%%%%%%%%%%%%%%%%%%%%%%%%%%%%%%%%%%%%%%%%%%%%%%%%%%%%
In the entrance channel, we describe the process of collision of a spherical projectile nucleus and a
deformed target nucleus.
In order to fix the shape of such a system (Fig.~\ref{4param}), it is necessary to use at least four
parameters (four collective coordinates). The $r$ pa\-ra\-me\-ter describes the distance between the centers of mass of colliding nuclei, the  $\alpha_t$ and $\alpha_p$ are parameters of quadrupole deformation of
interacting nuclei and the orientation parameter $\theta_t$ defines as the angle between the symmetry axis of
the deformed target nucleus and the line connecting centers of mass of the colliding nuclei.
Thus, it is assumed that the deformation of each of colliding nuclei can be described by only one parameter.

The potential energy of the system in the entrance channel includes the energy of the Coulomb and nuclear interactions \cite{Gross:78, Frobrich:84}, its rotational energy \cite{Kosenko:2008}, as well as the deformation energy of each nuclei,
\begin{eqnarray}
V^{I}_{\rm pot}&=&V_{\rm Coul}+V_{\rm GK}+E^{I}_{\rm rot}+E_{\rm def}^{(t)}+E_{\rm def}^{(p)}. \label{eq:potI}
\end{eqnarray}

\begin{figure}[htp]
\includegraphics[width=0.48\textwidth]{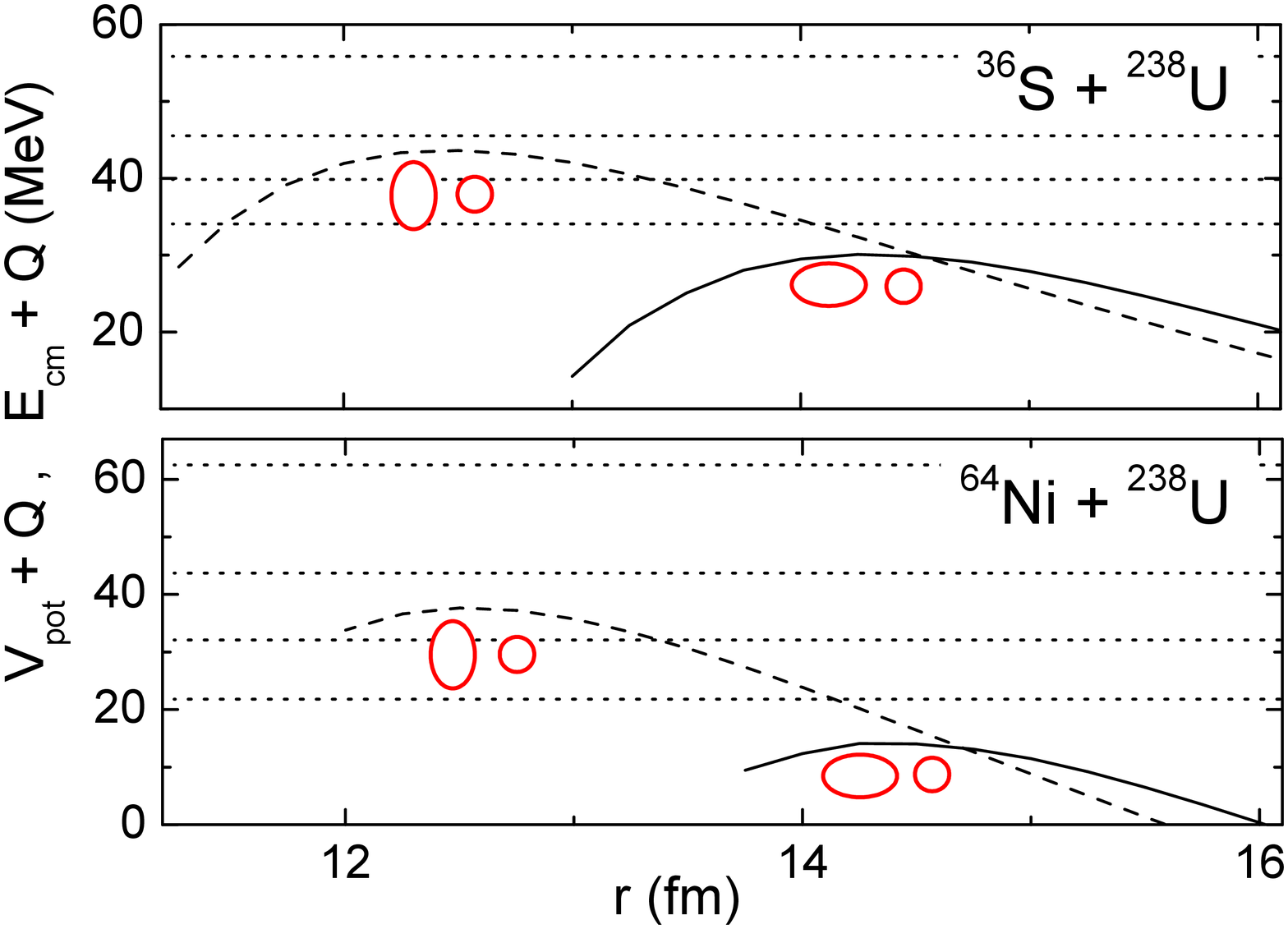}
\caption{The potential energy $V^{I}_{pot}$ \protect\req{eq:potI} of colliding ions ${\rm ^{36}S + ^{238}U}$ and ${\rm ^{64}Ni + ^{238}U}$ in the fusion channel for $L=0$, $\theta_t=0$ (solid) and $\theta_t=90^{\rm o}$ (dashed). Dotted horizontal lines are the reaction energy $E_x=E_{\rm cm}+Q$}
\label{barrier}
\end{figure}

The dependence of the potential energy of the system on the parameter $r$ is shown in Fig.~\ref{barrier}.
The dotted horizontal lines in this Figure are the reaction energies $E_x$=57.7, 47.3, 41.6, 35.8 MeV for the ${\rm ^{36}S + ^{238}U}$ reaction and $E_x$=64.1, 45.1, 33.5, 23.2 MeV for the ${\rm ^{64}Ni + ^{238}U}$ reaction, at which the fission and quasifission of composite systems with $Z$=108, 120 were investigated in \cite{Kozulin:2016}. 
 From this figure it is clear that the height of the Coulomb barrier depends very much on the orientation of the target nucleus.
%%%%%%%%%%%%%%%%%%%%%%%%%%%%%%%%%%%%%%%%%%%%%%%%%%%%%%%%
%%%%%%%%%%%%%%%%%%%%%%%%%%%%%%%%%%%%%%%%%%%%%%%%%%%%%%%%

The initial value of $r$  is chosen by the requirement that the nuclear interaction can be neglected and the Coulomb interaction between ions does not depend on their deformations and mutual orientation.
The value $r=50$~fm used in this paper satisfies these criteria well. The initial orientation of the target nucleus is distributed randomly, and the initial shape of the target nucleus corresponds to its ground state. At the initial moment of time, the movement in the system occurs only along the radial coordinate.

Starting with the initial value of collective variables and solving equations of motion (\ref{eq:UL}), one can determine the shape parameters of the system and the corresponding momentum at the next moments of time.

The exchange of energy between the collective and the
single-par\-ti\-cle degrees of freedom in the system being
considered is induced both by the relative motion of
colliding nuclei and by their deformation. Quantitatively,
the exchange of energy is characterized
by the inverse mass ($\mu=1/m$) and friction tensors.

The deformation of target and projectile are determined by one parameter $\alpha_t$ or $\alpha_p$. Thus, all four collective parameters in the entrance channel are "orthogonal"\  to each other, i.e. the mass and inverse mass tensors are diagonal. The diagonal components of the mass tensor describe the inertia of the system with respect to the motion along the corresponding degrees of freedom, namely inertia of the system with respect to the radial motion  is described by its reduced mass $M$, inertia of the system with respect to the deformation of each of the nucleus is described by mass $m_{\alpha \beta}^{I}$ tensors of isolated deformed nucleus (they were specified above in Eq. (\ref{masszero})), inertia of the system with respect to the rotation of the deformed
target nucleus is described by its rigid-body moment of inertia $J_{\rm t}$ arbitrarily oriented in space.

To determine components of the friction tensor we use equation: 
\bel{fric-one}
\gamma_{\beta \nu}=\gamma_{\beta \nu}^{\rm fus}+\delta_{\beta \alpha_t}\delta_{\alpha_t\nu}\gamma_{\alpha_t \alpha_t}^{I}+\delta_{\beta \alpha_p}\delta_{\alpha_p\nu}\gamma_{\alpha_p \alpha_p}^{I}.
\end{equation} 
The first term $\gamma_{\beta \nu}^{\rm fus}$ in this equation is determined in accordance with the surface-friction model \cite{Frobrich:84}. It depends on relative motion of the colliding nuclei. 
Second and third terms are components of the friction $\gamma_{\alpha \beta}^{I}$ tensor of isolated deformed target and projectile nuclei (specified above in Eqs. (\ref{friczero})). So, in the same way, as it was done in \cite{LVL:75:2012}, diagonal components of the friction tensor responsible for energy dissipation during the deformation of each of the nuclei are summed with the corresponding components obtained in the linear response theory.

%\begin{eqnarray}\label{fus}
%\mu_{\beta \nu}&=&\delta_{\beta r}\delta_{r \nu}M+\delta_{\beta \theta}\delta_{\theta \nu}J_{\rm t}+\nonumber\\
%&+&\delta_{\beta \alpha_t}\delta_{\alpha_t \nu}\mu_{\alpha_t \alpha_t}^{\rm nucl}+\delta_{\beta \alpha_p}\delta_{\alpha_p \nu}\mu_{\alpha_p \alpha_p}^{\rm nucl}\\
%\gamma_{\beta \nu}&=&\gamma_{\beta \nu}^{\rm fus}+\delta_{\beta \alpha_t}\delta_{\alpha_t\nu}\gamma_{\alpha_t \alpha_t}^{\rm nucl}+\delta_{\beta \alpha_p}\delta_{\alpha_p\nu}\gamma_{\alpha_p \alpha_p}^{\rm nucl}
%\end{eqnarray}
%The friction $\gamma_{\alpha \beta}^{\rm nucl}$ and mass $m_{\alpha \beta}^{\rm nucl}$ tensors of isolated deformed nucleus were specified above in Eqs. (\ref{friczero})) and (\ref{masszero}). Here we have denoted by $M$ the reduced mass of the system and by $J_{\rm t}$ rigid-body moment of inertia of the deformed target nucleus arbitrarily oriented in space. The deformation of target and projectile are fixed by one parameter $\alpha_t$ or $\alpha_p$. Thus, all four collective parameters in the entrance channel are "orthogonal"\  to each other, i.e. the mass and inverse mass tensors \req{fus} are diagonal. The components of the friction tensor $\gamma_{\beta \nu}^{\rm fus}$ are determined in accordance with the surface-friction model \cite{Frobrich:84}. In the same way as it was done in \cite{LVL:75:2012}, its diagonal components responsible for energy dissipation during deformation of each of the nuclei are summed with the corresponding components obtained in the linear response theory.

Due to the presence in the Langevin equations of the random force term, starting the calculation from the same point in the space of deformation parameters, one can get an infinitely large number of possible variants of the evolution of the system (similar to the trajectories of the Brownian particle in the space of collective co-\-ordinates describing the state of the system).

For the fixed value of the angular momentum of the system $L$, the heights of Coulomb barriers will be different for different trajectories. Part of the trajectories will be reflected by the Coulomb barrier. Such events correspond to the deep inelastic collisions. Part of the trajectories $N_{\rm bar} (L)$ will overcome the barrier. Knowing the initial number of trajectories $N(L)$ with angular momentum $L$, we can find the probability and cross sections (partial $\sigma_ {\rm bar} (L)$ and full $\sigma_{\rm bar}$) of crossing the Coulomb barrier:
\begin{eqnarray}\label{sigma}
P_{\rm bar} (L)= N_{\rm bar} (L)/N(L);\nonumber\\
\sigma_{\rm bar} (L)= (\pi /k^2) (2L+1) P_{\rm bar} (L);\\\label{eq:Sig_bar}
\sigma_{\rm bar}=\sum _L {\sigma_{\rm bar} (L)}\nonumber,
\end{eqnarray}
where $k^2$ is given by $k^2=2 M E_{cm}/\hbar^2$ with $M$ being the reduced mass in the entrance channel and $E_{cm}$ - the incident energy in the center-of-mass frame.
The first stage calculations are stopped at the moment when the system passes through the Coulomb barrier,  or reaches the internal turning point for the subbarrier fusion.
The values of the deformation parameters of the system, as well as the values of potential, kinetic and internal energy, are recorded. So, the distance between the centers of mass of the colliding nuclei $r$ depends on the point at which the system crossed the Coulomb barrier. With this information, we begin to describe the evolution of a highly deformed system formed after touching of the initial nuclei.
%%%%%%%%%%%%%%%%%%%%%%%%%%%%%%%%%%%%%%%%%%%%%%%%%%%%%%%%%%%%%%%%%%%%%
\subsection{Transition procedure}
\begin{figure}[htp]
\includegraphics[width=0.4\textwidth]{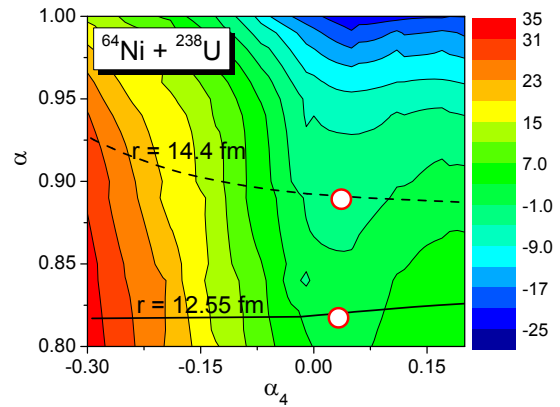}
\caption{The potential energy $E_{\rm def}$ of the combined system formed in the reaction ${\rm ^{64}Ni + ^{238}U}$ as a function of the parameters $\alpha_4, \alpha$ ($\alpha_1$ is fixed by the mass asymmetry of ${\rm ^{64}Ni + ^{238}U}$) system. The dashed line corresponds to fixed distance $r=14.4$ fm between the centers of mass of ions for the nose to nose touching configuration, and the solid line ($r=12.55$) corresponds to side to side configuration (see Fig.~\protect\ref{barrier}).  The circle marks the point where potential energy is minimal with respect to variations of $\alpha_4$.}
\label{NiUPot}
\end{figure}

%%%%%%%%%%%%%%%%%%%%%%%%%%%%%%%%%%%%%%%%%%%%%%%%%%%%%%%%%%%%%%%%%%%%%
The system formed after the touching of colliding nuclei is a highly deformed mass-asymmetric system with a well-pronounced neck.
To describe the shape of such systems, one needs to introduce at least three parameters that are responsible for the thickness of the neck, the distribution of the mass relative to the neck, and the elongation of the entire system. In the used in present work shape parametrization based on Cassini ovaloids, we consider three deformation parameters $\alpha, \alpha_1, \alpha_4$ that regulate the total elongation, the mass asymmetry and the neck radius, correspondingly.
The two of these pa\-ra\-me\-ters ($\alpha, \alpha_1$) can be found from the requirement that elongation and the mass asymmetry of the compact system is the same as that of two ions at the touching point. Unfortunately, the neck parameter for the touching system is not defined. So, it was assumed in \cite{LVL:2019} that the compact system attains the shape that corresponds to the minimum of deformation energy with respect to $\alpha_4$ (for given $\alpha$ and $\alpha_1$). The demonstration of the definition of $\alpha_4$ by such procedure is presented in Fig.~\ref{NiUPot}.

%%%%%%%%%%%%%%%%%%%%%%%%%%%%%%%%%%%%%%%%%%%%%%%%%%%%%%%%%%%%%%%%%%%%%%%%%%%%%%%%%%%%%%%%%%%
\subsection{The evolution of combined system}
%%%%%%%%%%%%%%%%%%%%%%%%%%%%%%%%%%%%%%%%%%%%%%%%%%%%%%%%%%%%%%%%%%%%%%%%%%%%%%%%%%%%%%%%%
After the initial parameters of the mono-system are set, we start solving the Langevin equations (\ref{eq:UL}).
The potential energy of the system included in these equations is the sum of deformation and rotation energies,
\begin{eqnarray}
V^{II}_{\rm pot}&=&E_{\rm def}+E^{II}_{\rm rot}. \label{eq:potII}
\end{eqnarray}

Tensors $\gamma_{\mu\nu}$ and $m_{\mu\nu}$ \cite{Ivanyuk:1999}, which were mentioned above, characterize completely the inertia and friction properties of the combined system.

After the start of calculations, all collective parameters of the system can change, directing it either to the ground state or to the scission line. The main change is however along with the mass asymmetric coordinate $\alpha_1$.   The outcome of Langevin equations depends very much on how much the mass asymmetric coordinate has changed before the fission.

If masses of separated parts of the system did not change much from the masses of colliding ions, then the deep inelastic collisions occur. If the masses change much, then such events correspond to fission or quasi-fission events. The latter differ from each other in how close the system came to the ground state before the separation occurred.

\begin{figure}[hb]
\includegraphics[width=0.48\textwidth]{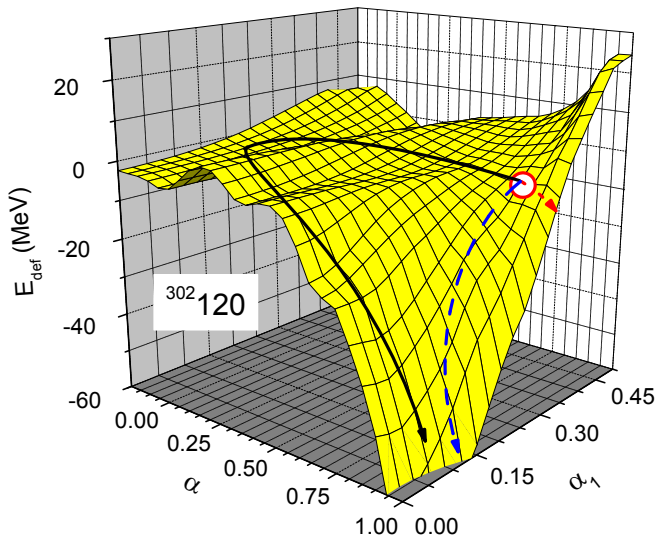}
\caption{The dependence of the potential energy of the system $^{302}120$ on the parameters $\alpha$ and $\alpha_1$ ($\alpha_4=0$). The circle shows the approximate position of the initial point of evolution of the mono-system formed in the reaction ${\rm ^{64}Ni + ^{238}U}$. The arrows show the possible directions of evolution of the mono-system: dot -- deep inelastic collision, dash -- quasi-fission, solid -- fission.}
\label{120Pot}
\end{figure}

\begin{figure}[ht]
\includegraphics[width=0.48\textwidth]{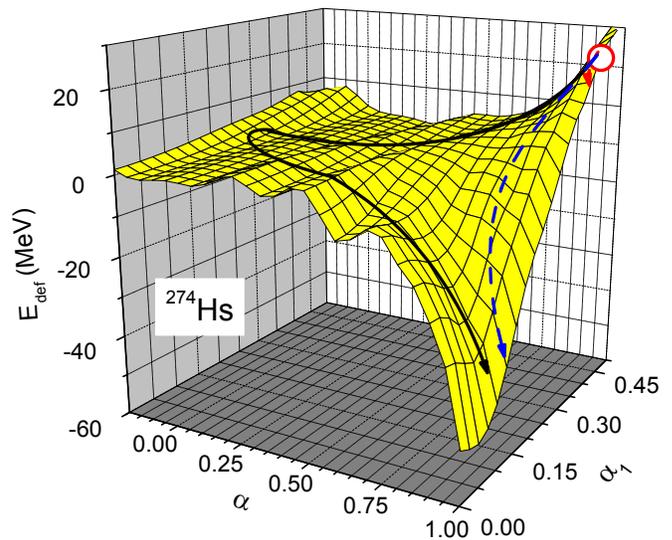}
\caption{The same as in Fig.~\ref{120Pot}, but for $^{274}$Hs nucleus, formed in the reaction ${\rm ^{36}S + ^{238}U}$.}\label{108Pot}
\end{figure}

In the Fig.~\ref{120Pot} and Fig.~\ref{108Pot} we show the dependence of deformation energy ($L=0$) of synthesized nuclei $^{302}120$ and $^{274}$Hs on the parameters $\alpha$ è $\alpha_1$ ($\alpha_4=0$). The initial deformation of the mono-system  for $^{302}120$ and $^{274}$Hs is marked by circles. Possible directions of its evolution are shown by arrows.  It is clearly seen that in case of $^{274}$Hs the system has more chances to come to the ground state compared with $^{302}120$.

During the evolution of combined system the total energy $E_{\rm tot}$ is shared between the local potential, kinetic and excitation energies
\bel{etotc}
E_{\rm tot}=V_{\rm pot}(q)+E_{\rm kin}(q)+E^*(q)
\end{equation}
Taking into account Eq.~\req{etota} for $E_{\rm tot}$ the local excitation energy is brought to the form
\bel{excc}
E^*(q)=E_x-(V_{\rm pot}(q)-E_{\rm gs}^{(t+p)})-E_{\rm kin}(q)
\end{equation}
Note, that the local excitation energy $E^*(q)$ does not coincide with  $E_x$. The probability of particles or $\gamma$-quanta emission and the kinetic energies of emitted particles is defined mainly by the local excitation energy $E^*(q)$.
With some probability, the system could also avoid fission and form the evaporation residue. This event is realized if the system being near the ground state will reduce its excitation energy by evaporating light particles (primarily neutrons) or emitting gamma-rays. The probability of these processes is estimated in the framework of the statistical model \cite{mebel:92} at each step of integration of Langevin equations (\ref{eq:UL}).
%%%%%%%%%%%%%%%%%%%%%%%%%%%%%%%%%%%%%%%%%%%%%%%%%%%%%%%%%%%%%%%%%%%%%%%%%%%%%%%%%%%
%%%%%%%%%%%%%%%%%%%%%%%%%%%%%%%%%%%%%%%%%%%%%%%%%%%%%%%%%%%%%%%%%%%%%%%%%%%%%%%%%%%%%%%%

We calculate the evolution of the
compact system either until it crosses the fission barrier back and splits into two fragments
or until it gets de-excited by the emission of light particles and gamma
rays and forms the evaporation residue.
%%%%%%%%%%%%%%%%%%%%%%%%%%%%%%%%%%%%%%%%%%%%%%%%%%%%%%%%%%%%%%%%%%%%%%%%%%%%%%%%%%
%%%%%%%%%%%%%%%%%%%%%%%%%%%%%%%%%%%%%%%%%%%%%%%%%%%%%%%%%%%%%%%%%%%%%%%%%%%%%%%%%%%%

\begin{figure}[ht]
\includegraphics[width=0.48\textwidth]{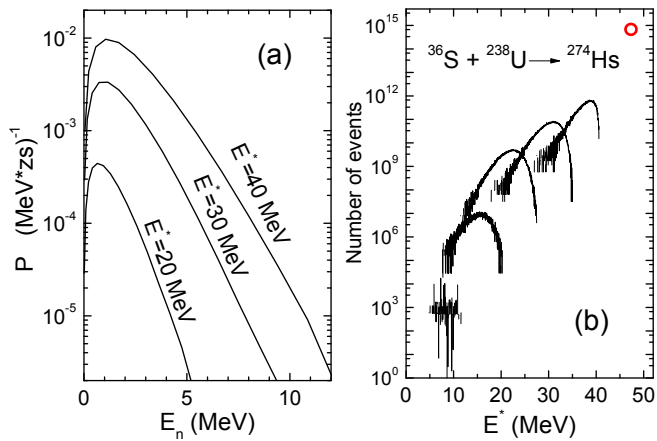}
\caption{(a) The probability $P(E_n)$ (Eq. \protect\ref{Pne}) of neutron emission per time unit as a function of neutron kinetic energy $E_n$ for few values of excitation energy $E^*$ of compound nucleus $^{274}$Hs; (b) The dependence of number of trajectories that stay at the ground state region ($\alpha\le$ 0.1) after emission of 1, 2, 3, 4 and 5 neutrons on the excitation energy $E^*$ of the compound nucleus. The initial excitation energy $E^*$=47.3 MeV is marked by the circle.}
\label{Udistr}
\end{figure}

In order to form the evaporation residue, the system should release the excitation energy by the evaporation of light particles and $\gamma$-quanta.  We describe the particle evaporation from an excited nucleus by the statistical method proposed in \cite{mebel:92}, see also \cite{Kosenko:2008}.
On each step of integration of Langevin equations by the hit-and-miss method, we check if the particle was emitted and what kind of partible was emitted.  The expressions for the  evaporation widths $\Gamma_j$ (j$\equiv n, p, d, t, ^3He, \alpha$) and $\Gamma_\gamma$ are given in \cite{mebel:92}.
In particular, for the probability $P_n$ of emitting neutron within the time step $\Delta t$ of integration of Langevin equations one can find
\bel{Pn}
P_n=\Delta t \int_{0}^{E^*_n-B_n}  P(E_n)\,dE_n,
\end{equation}
where $P(E_n)$ is the probability  of emitting neutron with a certain energy $E_n$ per time unit,
\bel{Pne}
P(E_n)=\frac{(2s_n+1) m_n}{\pi^2 \hbar^3\rho_0(E^*_0)} \sigma_{inv}(E_n)\,E_n\,\rho_n(E^*_n-B_n-E_n).
\end{equation}
Here, $\rho_0$ and $\rho_n$ are the level densities in the primary nucleus and the nucleus formed
after the neutron emission; $s_n$, $m_n$, $B_n$ are
the spin of the emitted neutron, its mass and its binding energy;
$\sigma_{inv}(E_n)$ is the cross section for the absorption of a neutron  with kinetic energy $E_n$ by the considered nucleus; $E^*_n=E^*-\Delta_n$; $E^*_0=E^*-\Delta_0$; $E^*$ is the compound-nucleus excitation energy; and $\Delta_n$ and $\Delta_0$ are the pairing gaps for the residual and the primary nucleus, respectively. The probability $P_n$ for $^{274}$Hs  nucleus is shown in Fig.~\ref{Udistr}(a).
%%%%%%%%%%%%%%%%%%%%%%%%%%%%%%%%%%%%%%%%%%%%%%%%%%%%%%%%%%%%%%%%%%%%%%%%%%%%%%%%%
%%%%%%%%%%%%%%%%%%%%%%%%%%%%%%%%%%%%%%%%%%%%%%%%%%%%%%%%%%%%%%%%%%%%%%%%%%%%%%%%%%%%%%

After finding the sum of probabilities to evaporate any particle (total probability), which is calculated in the same way as it was demonstrated here for neutrons (\ref{Pn}),  by the hit-and-miss method we determine which particle, if any, was evaporated. For this, we generate a random number $\xi$ between zero and unity and compare it with total probability. If this random number is smaller than
the total probability, it is assumed that a particle is emitted at the current step of solving Langevin equations. The kind of a
particle is determined again at random proportionally to the known probability of evaporation of any particle.
Then, knowing the dependence of the particle evaporation probability on its kinetic energy, we again randomly choose its kinetic energy. As one can see from Fig.~\ref{Udistr}(a), the most probable kinetic energy of the evaporated neutrons is close to 1-2 MeV.

If some particle is emitted, the
binding energy of this particle is subtracted from the
excitation energy of the system, the deformation energy, and the transport coefficient are replaced by these for smaller particle number.  The calculations show that at high value of $E_x$, $E_x\approx 50$ MeV up to 4-5 neutrons can be emitted.

During the evolution of the system from the touching configuration, it has a very high probability to split into two pieces and form the product of quasi-fission. A very few trajectories would reach the ground state deformation. Some of them could decrease their excitation energy by light particles or gamma emission. The dependence of the probability to come to the ground state on the number of evaporated neutrons will be discussed in the next section. Here we will illustrate the  deexcitation process and evaporation residue formation in the reaction ${\rm ^{36}S + ^{238}U}\to^{274}$Hs for the case when trajectories come to the ground state without evaporation of any particles with their initial energy $E_x$=47.3 MeV. Fig.~\ref{Udistr}(b) demonstrates the deexcitation process. The "survived" nuclei could reduce the excitation energy by the first evaporation of neutron. Since the kinetic energy of first emitted neutron is not fixed but distributed around some most probable value, see Fig.~\ref{Udistr}(a), after neutron emission one gets the distribution of events around most probable excitation energy $E^*$=38.7 MeV (first peak on the right in Fig.~\ref{Udistr}(b)). The excitation energy after evaporation of the first neutron is still high, the main part of nuclei would fission,  the rest would emit the second neutron and form the second peak on the right in Fig.~\ref{Udistr}(b) with the most probable excitation energy $E^*$=31.0 MeV. The process of fission and neutron emission would continue until the excitation energy becomes smaller than the fission barrier. In this case, one can say that the evaporation residue was formed. The number of trajectories that formed the evaporation residue in case of $^{274}$Hs is by 13-15 orders of magnitude smaller than the initial number of trajectories, that reached the ground state.

Knowing probability of the system formed after collision of the initial nuclei to form the compound nucleus (fusion process) $P_{\rm CN}(L)$ and probability for the compound nuclear to survive against fission $W_{\rm sur}(L)$ one can calculate fusion $\sigma_{\rm fus}$ and evaporation residue formation $\sigma_{\rm er}$ cross sections:
\bel{sfus}
\sigma_{\rm fus}=\sum_L \sigma_{\rm fus}(L)=\sum_L \sigma_{\rm bar}(L) P_{\rm CN} (L)
\end{equation} 
and 
\bel{ser}
\sigma_{\rm er}=\sum_L \sigma_{\rm er}(L)=\sum_L \sigma_{\rm bar}(L) P_{\rm CN}(L) W_{\rm sur} (L),
\end{equation} 
where $\sigma_{\rm fus}(L)$ and $\sigma_{\rm er}(L)$ are fusion and evaporation residue formation partial cross sections. 
%%%%%%%%%%%%%%%%%%%%%%%%%%%%%%%%%%%%%%%%%%%%%%%%%%%%%%%%%%%%%%%%%%%%%%%%%%%%%%%%%%%
\section{Results and discussions}
%%%%%%%%%%%%%%%%%%%%%%%%%%%%%%%%%%%%%%%%%%%%%%%%%%%%%%%%%%%%%%%%%%%%%%%%%%%%%%%%%%%%%%
In present work we  consider the fusion-fission process in reactions ${\rm ^{36}S + ^{238}U}\to^{274}$Hs  and $^{64}$Ni + $^{238}$U $\to$ $^{302}$120. The calculations of the entrance channel provide for these reactions the Coulomb barrier penetration cross sections. Their values should be close to the values of the capture cross sections, obtained in the experiments. It should be noted that the probability of capture is determined by the probability that fission or quasi-fission events will occur during the reaction. And it does not include the probability of a deep inelastic scattering process, which, in principle, can occur at the second stage of the reaction. Therefore, the cross-sections of the system crossing the Coulomb barrier obtained at the end of the first stage of calculation may be slightly larger than the capture cross sections.

\begin{figure}[htp]
\includegraphics[width=0.48\textwidth]{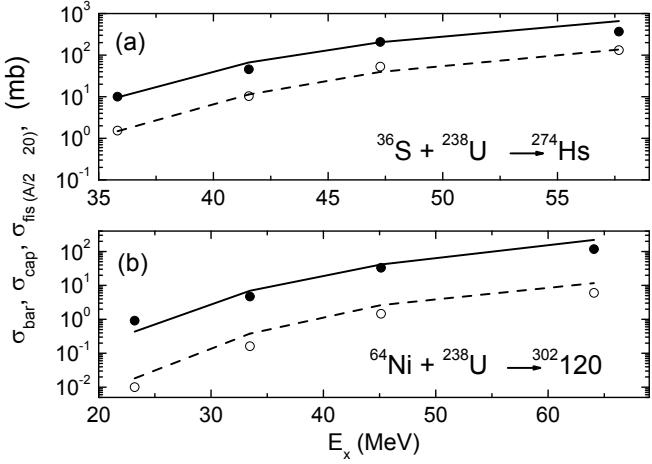}
\caption{The Coulomb barrier penetration cross sections (solid line) for the  reaction ${\rm ^{36}S + ^{238}U}$ (a)  and ${\rm ^{64}Ni + ^{238}U}$ (b) as function the reaction energy $E_x=E_{c.m.}+ Q$. Closed circles are the experimental data on capture cross sections \cite{Kozulin:2016}. The cross sections of the fission of the compact system, corresponding to the fragments mass asymmetry $A/2\pm 20$ are shown by dash
 line (calculations) and  open circles (experimental data \cite{Kozulin:2016}).  }
\label{Cap}
\end{figure}
In Fig.~\ref{Cap} the cross sections of the Coulomb barrier penetration, the cross sections of almost symmetric (with the ratio of the masses of fragments $A/2\pm 20$) fission and quasi-fission of the system, formed after touching of the initial nuclei, are given. For comparison, the corresponding experimental data \cite{Kozulin:2016} are also presented. It can be seen that the theoretical calculations agree rather well with the experimental data.

%%%%%%%%%%%%%%%%%%%%%%%%%%%%%%%%%%%%%%%%%%%%%%%%%%%%%%%%%%%%%%%%%%%%%%%%%%%%%%%%%%%%%%%%
\begin{figure}[htp]
\includegraphics[width=0.48\textwidth]{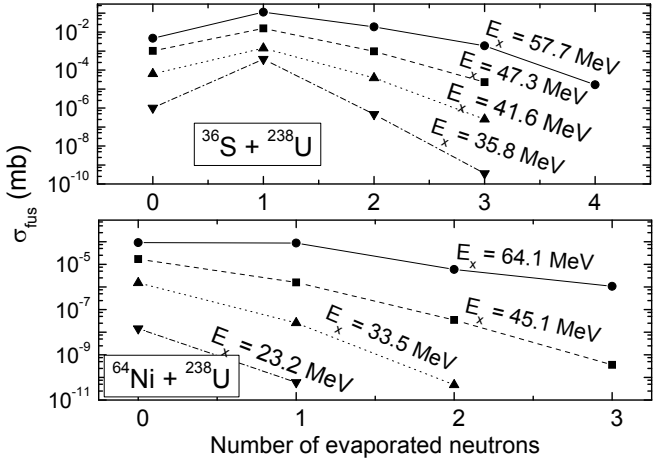}
\caption{The dependence of fusion cross section, obtained for the reaction $^{36}$S+$^{238}$U $\to^{274-x}$Hs+$xn$ and
$^{64}$Ni + $^{238}$U$\to$ $^{302-x}$120 + $xn$,  on the number of emitted neutrons.}
\label{GSNiUsig}
\end{figure}
%%%%%%%%%%%%%%%%%%%%%%%%%%%%%%%%%%%%%%%%%%%%%%%%%%%%%%%%%%%%%%%%%%%%%%%%%%%%%%%%%%%%%%%%%%
The small fraction of trajectories do not undergo quasi-fission immediately and can reach the region of the ground state deformation. Such trajectories can be considered as leading to the fusion of the colliding ions. During the further evolution, the mono-system can evaporate few neutrons or gamma rays. So, the compound nucleus will be a set of different isotopes with different excitation energies.

The values of fusion cross sections (events reaching the ground state), for all considered energies of reaction ${\rm ^{64}Ni + ^{238}U} \to ^{302-x}120 + xn$ are given in Fig.~\ref{GSNiUsig}.

\begin{figure}[ht]
\includegraphics[width=0.48\textwidth]{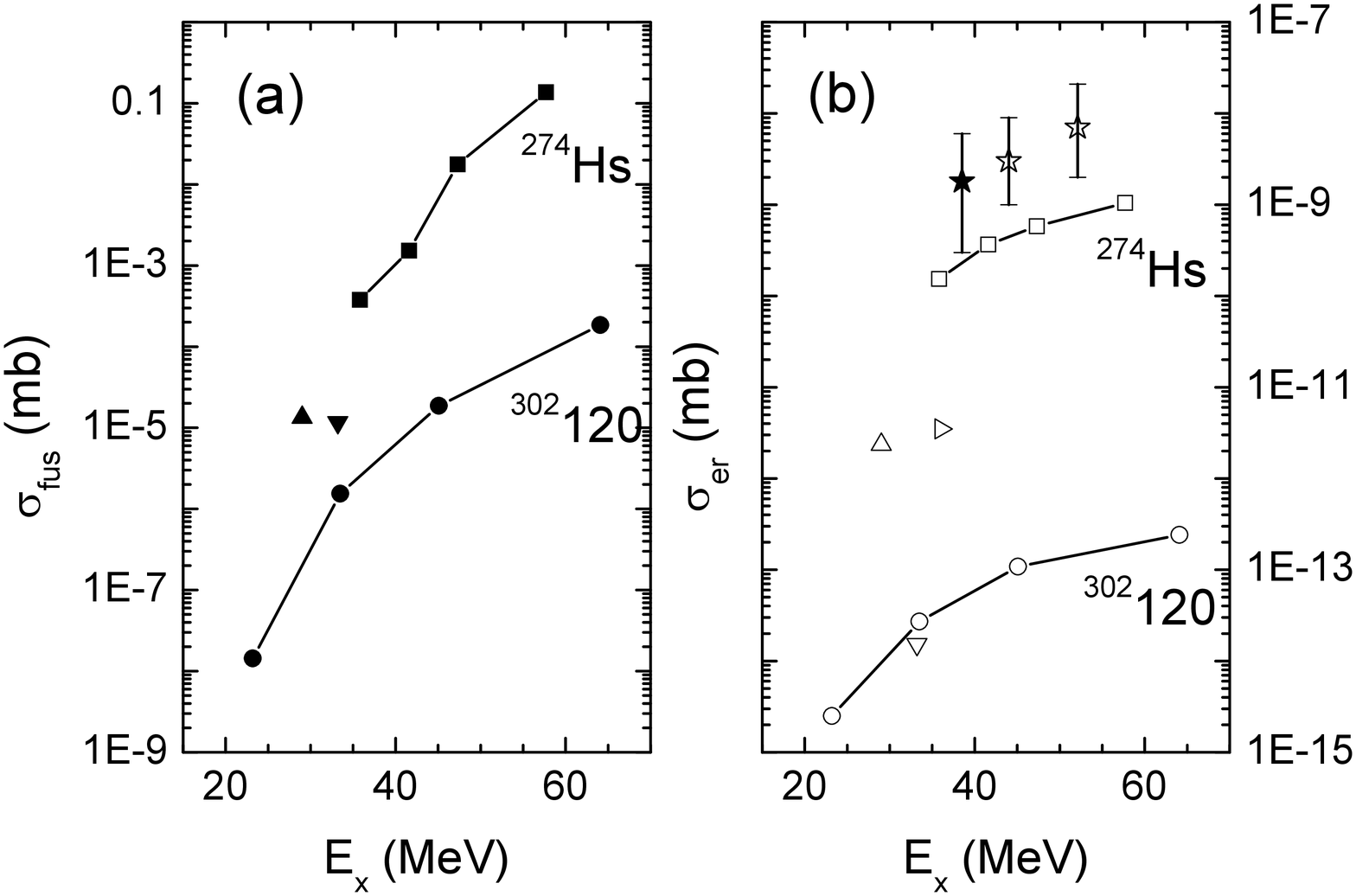}
\caption{(a) The dependence of the total fusion cross-sections of the isotopes obtained in the reactions $^{36}$S+$^{238}$U $\to^{274}$Hs and ${\rm ^{64}Ni + ^{238}U}\to ^{302}$120 on the initial excitation energy $E_x$. The triangles are the results from \protect\cite{Hagino:19}; (b) The evaporation residues formation cross-sections for the same reactions. The up and down triangles mark the data obtained in \cite{Hagino:19}. The backside triangle and the stars are the results from \protect\cite{Zagrebaev:08} and  \cite{Nishio:10,Dvorak:06}.}
\label{GSERNiU}
\end{figure}

The obtained results for the fusion cross-sections and the excitation energies of the corresponding isotopes can be used for calculation of the evaporation residues formation cross-sections. The summed over all isotopes values of the fusion cross sections and the evaporation residues formation cross sections are given in Fig.~\ref{GSERNiU}.

The first superheavy element with $Z$=108, $^{266}$Hs, was synthesized at GSI, Darmstadt \cite{Muenzen:84} in the so-called cold fusion reaction $^{58}$Fe + $^{208}$Pb $\to^{266}$Hs with the doubly magic $^{208}$Pb as a target. The excitation energy of compound system in this reaction was rather low $18\pm 2$ MeV and only one neutron was emitted during the de-excitation process.  For the three observed events, the production cross section $\sigma_{\rm er}=19\pm^{18}_{11}$ pb was deduced.

The heavier superheavies $Z$=114-118 were produced at JINR, Dubna in the so-called warm fusion reactions. In these reactions, the initial excitation of the compound nucleus was of the order of $(30\sim 40)$ MeV, consequently, up to 4-5 neutrons were emitted and the residue formation cross section was much lower as compared with the cold fusion reactions. For the comparison of our calculated results for $^{274}$Hs we choose the available experimental results from similar reactions $^{34}$S+$^{238}$U$\to^{272}$Hs at $E_x$=38.5 MeV \cite{Nishio:10} and $^{26}$Mg+$^{248}$Cm$\to^{270}$Hs, at $E_x$=44 MeV and $E_x$=52.1 MeV \cite{Dvorak:06}. The last reaction is more mass-asymmetric than calculated here, so the higher values of $\sigma_{er}$ than ours  should be expected.

As one can see from the top part of Fig.~\ref{GSERNiU}(b) both experimental and calculated results grow with the increasing excitation energy $E_x$. The calculated results for $^{36}$S+$^{238}$U$\to^{274}$Hs reaction are on average by one order of magnitude smaller than the experimental cross sections from mentioned above reactions. Taking into account the uncertainty of experimental results, the discrepancy between theory and experiment is not so large.

The calculated data for ${\rm ^{64}Ni + ^{238}U}\to ^{302}$120 reaction are shown in the bottom part of Fig.~\ref{GSERNiU}. As one could expect, the fusion cross-section for $^{302}$120 is by few orders of magnitude smaller as compared with that of $^{274}$Hs.
Consequently, the evaporation residues formation cross section $\sigma_{er}$ for $^{302}$120 is also much smaller as compared with that of $^{274}$Hs.
%%%%%%%%%%%%%%%%%%%%%%%%%%%%%%%%%%%%%%%%%%%%%%%%%%%%%%%%%%%%%%%%%%%%%%%%%%%%%%%%%%
%%%%%%%%%%%%%%%%%%%%%%%%%%%%%%%%%%%%%%%%%%%%%%%%%%%%%%%%%%%%%%%%%%%%%%%%%%%%%%%%%%%%

For the comparison we show the results of time-dependent Hartree-Fock plus Langevin approach for hot fusion reactions \cite{Hagino:19} for more mass-asymmetric combinations of the target and projectile, $^{254}$Fm + $^{48}$Ca, $P_{\rm CN}W_{\rm sur}$=302*10$^{-13}$  at $E_x$=29.0 MeV (up-triangle in Fig.~\ref{GSERNiU}(b)) and
$^{248}$Cm + $^{54}$Cr, $P_{\rm CN}W_{\rm sur}$=2.47*10$^{-13}$ at $E_x$=33.2 MeV (down-triangle). In order to bring the probabilities shown in Table 1 of \cite{Hagino:19} to the same dimension as our calculated cross sections we have multiplied the probabilities of \cite{Hagino:19} by the factor $\pi/k^2$, see Eq.\req{sigma}. Unfortunately, in \cite{Hagino:19} the results of calculations are presented only for the case $L=0$,  one term in the sum (see Eq.\req{ser}). The account of higher orbital momenta should increase the value of this sum. %The account of higher orbital momenta may increase substantially the calculated probability of compound nucleus formation. 
Thus, the calculations within the model of \cite{Hagino:19} for higher orbital momenta are very much desirable.

The backside triangle shows the evaporation residue cross section calculated for the reaction $^{64}$Ni + $^{238}$U $\to^{302}$120  at $E_x$=36 MeV in dynamical (up to compound nucleus formation) statistical (survival probability calculations) model \cite{Zagrebaev:08}. Our calculated results (open circles in Fig.~\ref{GSERNiU}(b)) are in the middle between the calculations of \cite{Hagino:19} and \cite{Zagrebaev:08}, what is quite reasonable.

%%%%%%%%%%%%%%%%%%%%%%%%%%%%%%%%%%%%%%%%%%%%%%%%%%%%%%%%%%%%%%%%%%%%%%%%%%%%%%%%
\section{Conclusions}
%%%%%%%%%%%%%%%%%%%%%%%%%%%%%%%%%%%%%%%%%%%%%%%%%%%%%%%%%%%%%%%%%%%%%%%%%%%%%%%%
In the present work, reactions that differ from each other by the ratio of the masses of colliding nuclei almost twice were studied. We have applied a dynamical approach to calculate the evolution of the system starting from the approaching of the colliding ions to each other and up to fission (quasi fission) of the system, formed after touching of the initial nuclei or up to the evaporation residue formation.  We have demonstrated that our two-stage stochastic model for fusion-fission reactions describes rather well the existing experimental data for the synthesis of Hs isotopes. Thus, the values of the fusion cross sections and the evaporation residues formation cross section obtained for the reaction ${\rm ^{64}Ni + ^{238}U} \to ^{302-x}120 + xn$ should be reliable. 
According to our results, the most favorable energy of $^{64}$Ni ions should be close to $E_{\rm c.m.}$=300 MeV.

These data can be used for further advancement to the region of superheavy elements.

\begin{acknowledgments}
One of us (V. L.) would like to express his gratitude to the
Research Laboratory for Nuclear Reactors, Tokyo Institute of Technology, for the hospitality during his stay in
Japan.
\end{acknowledgments}

\end{document}